\documentclass[%
 reprint,
nofootinbib,
amsmath,amssymb,
prd,
]{revtex4-2}

\usepackage{graphicx}
\usepackage{dcolumn}
\usepackage{bm}
\usepackage{hyperref}

\def\tsp{\hspace{0.05em}}
\def\arXiv#1{\href{http://arxiv.org/abs/#1}{arXiv:#1}}
\def\arXiv#1#2{\href{http://arxiv.org/abs/#1}{arXiv:#1}}
\def\arXivid#1#2{\href{http://arxiv.org/abs/#1/#2}{#1/#2}}
\def\doi#1#2{\href{http://doi.org/#1}{#2}}

\begin{document}


\title{Analytic AC conductivities from holography}

\author{Jie Ren}
\author{Wenni Zheng}
\affiliation{School of Physics, Sun Yat-sen University, Guangzhou, 510275, China}

\date{\today}

\begin{abstract}
We find exact, analytic solutions of the holographic AC conductivity at arbitrary frequency $\omega$ and temperature $T$, in contrast to previous works where the AC conductivity was analytically obtained usually at small $\omega$ and $T$. These solutions enable us to study the analyticity properties of the current-current correlator $G(\omega)$ in detail. The first system we study is the AdS$_5$ planar black hole with momentum dissipation, whose extremal limit has an AdS$_2$ factor. Then we study AdS$_4$ and AdS$_5$ Einstein-dilaton systems whose special cases are maximal gauged supergravities. The solutions show how the poles move and how branch cuts emerge as the temperature varies. As a by-product, we obtain an R-current correlator in $\mathcal{N}=4$ super-Yang-Mills theory on a sphere at finite temperature in the large $N$ and strong coupling limit.
\end{abstract}

\maketitle

\tableofcontents

\section{Introduction}
The AdS/CFT correspondence \cite{Maldacena:1997re,Gubser:1998bc,Witten:1998qj} provides a powerful tool to study strongly interacting quantum systems,
 where the perturbative method does not work. An example is condensed matter systems without quasiparticles \cite{Hartnoll:2016apf,Hartnoll:2008vx,Faulkner:2009wj}. A key observable to calculate is the electrical conductivity, which is related to the current-current correlation function by the Kubo formula. Analyticity properties of correlation functions are of great interest to theoretical physicists. There have been a large number of works on calculating the conductivity. Specifically, the DC conductivity can often be calculated analytically.

However, analytic solutions of the AC conductivity are rare. Known solutions are usually from constant curvature spacetimes, such as pure AdS or the BTZ black hole \cite{Banados:1992wn,Ren:2010ha}. Another known solution is from the planar AdS$_5$ black hole, corresponding to the R-current correlator of the $\mathcal{N}=4$ super-Yang-Mills (SYM) theory \cite{Kovtun:2005ev,Myers:2007we}. However, in this system, all finite temperatures are equivalent. It is desirable to obtain analytic solutions of the AC conductivity that has a nontrivial temperature dependence. The analyticity properties of the conductivity can be studied as the temperature varies and the extremal limit is approached.

Momentum dissipation exists in real-world materials. Without any momentum dissipation, the DC conductivity for finite density systems is divergent. A simple way to introduce momentum dissipation is by adding axions (massless scalar fields) in the system \cite{Andrade:2013gsa,Davison:2014lua,Baggioli:2021xuv}. We consider zero density systems for simplicity. Since the momentum dissipation introduces another scale, the system has a nontrivial temperature dependence for the planar black hole. In the extremal limit, the IR geometry may have an AdS$_2$ factor or be a hyperscaling-violating geometry.

In this paper, we add momentum dissipation in zero density systems that are closely related to solutions of maximal gauged supergravities. Zero density matter without quasiparticles arises at some quantum critical points in condensed matter systems \cite{Hartnoll:2016apf}. One example is graphene, or electrons on a honeycomb lattice at a density of one electron per site. Another example is ultracold atoms in optical lattices with one atom for each minimum of the periodic potential. Zero density systems are particle-hole symmetric, and have electrical conductivity due to pair production. The gravity dual to zero density systems is neutral black holes.

We obtain a number of analytic solutions of the AC conductivity in an AdS$_5$ system and dilatonic AdS$_4$ and AdS$_5$ systems. Especially, the solutions at zero temperature clearly reveal branch cuts, delta functions, and gaps in certain regions of parameter space, and these features are challenging to infer numerically. Analyticity properties of the retarded current-current correlator $G_R(\omega)$ are studied. At finite temperature, there are poles in the lower-half complex $\omega$-plane. When the extremal limit is taken, the poles will become branch cuts via different moving patterns. Finally, as a by-product, an R-current correlator in SYM theory on a sphere at finite temperature is analytically calculated.

In Sec.~\ref{sec:emda}, we describe the model in a general framework. In Sec.~\ref{sec:schw5}, we obtain an analytic solution of the AC conductivity for the planar Schwarzschild-AdS$_5$ black hole with momentum dissipation. In Sec.~\ref{sec:ads4}, we consider a one-parameter family of AdS$_4$ systems at both zero and finite temperatures. In Sec.~\ref{sec:ads5}, we consider a one-parameter family of AdS$_5$ systems. In Sec.~\ref{sec:R-current}, we obtain an R-current correlator in SYM theory. In Sec.~\ref{sec:sum}, we conclude.

\section{Einstein-dilaton-axion systems}
\label{sec:emda}
We study zero density systems dual to AdS$_{d+1}$ with momentum dissipation.
The background geometry is determined by Einstein gravity, a neutral scalar $\phi$, and axions $\chi_i$ ($i=1,\cdots,d-1$). The holographic conductivity is calculated by perturbing the system with a Maxwell field $A_\mu$.
The Lagrangian density is \cite{Gouteraux:2014hca,Kiritsis:2015oxa}
\begin{equation}
\mathcal{L}=R-\frac14 Z(\phi)F^2-\frac12(\partial\phi)^2-V(\phi)
-\frac{1}{2}\sum_{i=1}^{d-1}(\partial\chi_i)^2,\label{eq:axionsd}
\end{equation}
where $\chi_i=a x_i$ satisfies the equations of motion and breaks the translation symmetry; $a$ is a parameter describing the strength of momentum dissipation. We consider solutions of neutral planar AdS black holes with the metric
\begin{equation}
  ds^2=-f(r)\tsp dt^2+f(r)^{-1}\tsp dr^2+U(r)\tsp d \vec{x}^2\, .\label{eq:metric}
\end{equation}

The conductivity as a function of frequency in strongly coupled systems dual to the AdS gravity can be calculated from the retarded current-current correlator by the AdS/CFT prescription.
To calculate the AC conductivity, we perturb the system by an alternating electric field along the $x\equiv x_1$ direction, which is achieved by adding a vector potential $\delta A_x(r,t)=e^{-i\omega t}A_x(r)$. 
The Maxwell's equations for the zero density system are given by
\begin{equation}
A_x''+\frac{(\sqrt{-g}\,Z(\phi)g^{rr}g^{xx})'}{\sqrt{-g}\,Z(\phi)g^{rr}g^{xx}}A_x'+\frac{\omega^2}{|g_{tt}|g^{rr}}A_x=0\,,
\label{eq:fluctuation}
\end{equation}
where the prime denotes derivative with respect to the radial coordinate $r$.
To obtain the retarded Green's function, we impose the infalling boundary condition at the horizon $r=r_h$: $A_x\propto(r-r_h)^{-i\omega/|f'(r_h)|}$.
At the AdS boundary $r\to\infty$, the leading and subleading terms of $A_x(r)$ are
\begin{equation}
A_x(r)=\mathcal{A}+\mathcal{B}\frac{L^{2(d-2)}}{r^{d-2}}+\mathcal{\tilde{B}}\frac{L^{2(d-2)}}{r^{d-2}} \log(\Lambda r)\,,
\label{eq:axsol}
\end{equation}
where $\Lambda$ is a cutoff scale taken to be 1. In AdS$_5$, $\mathcal{\tilde{B}}=\mathcal{A}\, \omega^2/2$, while in AdS$_4$, $\mathcal{\tilde{B}}=0$.
According to the AdS/CFT prescription \cite{Gubser:1998bc,Witten:1998qj,Son:2002sd}, $\mathcal{A}$ and $\mathcal{B}$ are the dual source of the current $\mathcal{O}$ and its expectation value $\langle{\mathcal{O}}\rangle$ in the boundary field theory, respectively.
Then the retarded Green's function corresponding to the current-current correlator $\langle J^xJ^x\rangle$ is \cite{Horowitz:2008bn}
 \begin{equation}
   G(\omega)=\begin{cases}
   \mathcal{B}/\mathcal{A}  &\text{for AdS$_4$}\,,\\
   2\,\mathcal{B}/\mathcal{A}-\omega^2/2 &\text{for AdS$_5$}\,.
   \end{cases}\label{eq:green}
 \end{equation}
By the Kubo formula, the conductivity is $\sigma(\omega)=G/i \omega$.

\section{Planar black hole in AdS$_5$}
\label{sec:schw5}
We start with the planar Schwarzschild-AdS$_5$ black hole with momentum dissipation. We take $V=-12/L^2$, $Z=1$ and $\phi=0$ in Eq.~\eqref{eq:axionsd}. The blackening factor of the metric with $U(r)=r^2$ is
\begin{equation}
  f(r)=\frac{1}{L^2}\biggl(r^2+r_h^2-\frac{a^2L^2}{4}\biggr)\biggl(1-\frac{r_h^2}{r^2}\biggr),
\label{eq:fsol}
\end{equation}
where $r_h$ is the horizon size. The Hawking temperature of this black hole is
\begin{equation}
  T=\frac{8r_h^2-a^2L^2}{8\pi r_h L^2}.
\end{equation}
The extremal limit is at $a=2\sqrt{2}r_h/L$, at which the IR geometry is AdS$_2\times\mathbb{R}^3$ \cite{Andrade:2013gsa}. As a comparison, the geometry at $T=0$ without axions is the pure AdS. Without loss of generality, we set $L=1$ and $r_h=1$.

The solution of the perturbation equation~\eqref{eq:fluctuation} with the infalling boundary condition at the horizon is
\begin{align}
  &A_x=\left(\frac{r^2-1}{r^2-1+2\pi T}\right)^{-\frac{i \omega }{4\pi T}} {_2F_1}\biggl(-\frac{i \omega\bigl(1-\sqrt{1-2\pi T}\bigr)}{4\pi T},\nonumber\\
  &\quad -\frac{i \omega(1+\sqrt{1-2\pi T})}{4\pi T};
  1-\frac{i \omega}{2\pi T};\frac{r^2-1}{r^2-1+2\pi T}\biggr),\label{eq:sol-AdS5}
\end{align}
where the parameter $a$ in the metric has been replaced by the temperature $T$.
To obtain the Green's function, we expand $A_x(r)$ at asymptotic infinity as Eq.~\eqref{eq:axsol}. By Eq.~\eqref{eq:green}, the Green's function is
\begin{align}
  &G(\omega)=i \omega-\frac{1}{2}\omega^2\biggl[
  \psi\biggl(1-\frac{i\omega\big(1-\sqrt{1-2\pi T}\big)}{4\pi T} \biggr)\nonumber\\
&\; +\psi\biggl(1-\frac{i \omega\big(1+\sqrt{1-2\pi T}\big)}{4\pi T}\bigg)+\log(2\pi T)+2\gamma\biggr],
\end{align}
where $\psi$ is the digamma function $\psi(z)=\Gamma'(z)/\Gamma(z)$ and $\gamma$ is Euler's constant. The poles of the Green's function are at
\begin{equation}
  \omega_n=-2 i n\bigl(1\pm\sqrt{1-2 \pi T}\bigr),\qquad n=1,2,3,\cdots\, .
\end{equation}

Interestingly, the real part of the conductivity can be expressed by an elementary function as 
\begin{multline}
\text{Re}[\sigma(\omega)]=\frac{\pi\omega}{4}\biggl(\coth\frac{\left(1-\sqrt{1-2 \pi  T}\right) \omega}{4 T}\\
+\coth\frac{\left(1+\sqrt{1-2 \pi  T}\right) \omega}{4 T}\biggr).
\end{multline}
From the indefinite integral over the real part of the conductivity, we obtain the following identity
\begin{equation}
  \int^{\infty}_{0} \Bigl(\text{Re}[\sigma(\omega)]-\frac{\pi\omega}{2}\Bigr)d\omega=\frac{\pi}{3}(1-\pi T)=\frac{\pi a^2}{24}.
\end{equation}
When $a=0$, i.e., without the momentum dissipation, the right-hand side is zero, which matches the sum rule in~\cite{Gulotta:2010cu}. This implies that the sum rule depends on the conservation of momentum.

\begin{figure*}
  \includegraphics[width=0.3\textwidth]{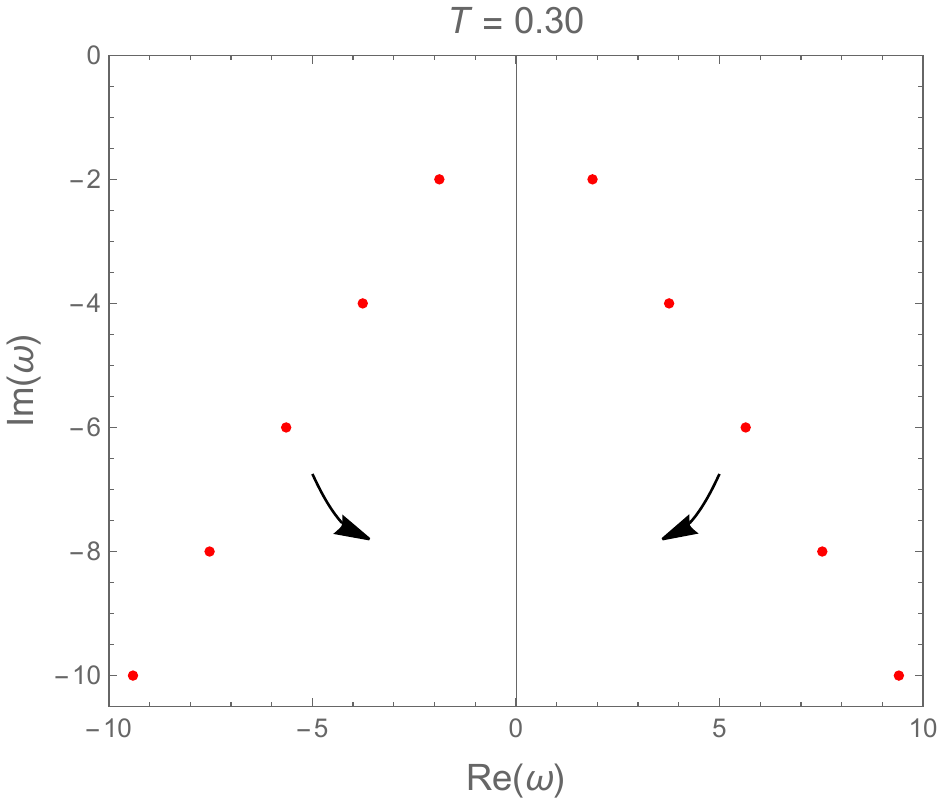}\qquad
  \includegraphics[width=0.3\textwidth]{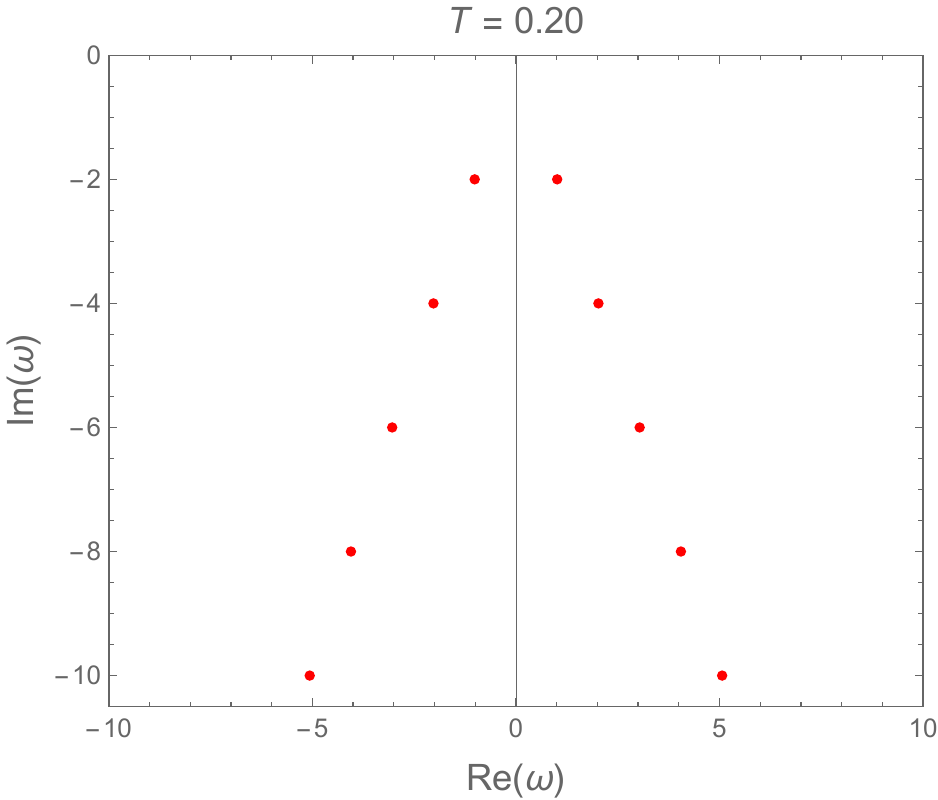}\\
  \vspace{5pt}
  \includegraphics[width=0.3\textwidth]{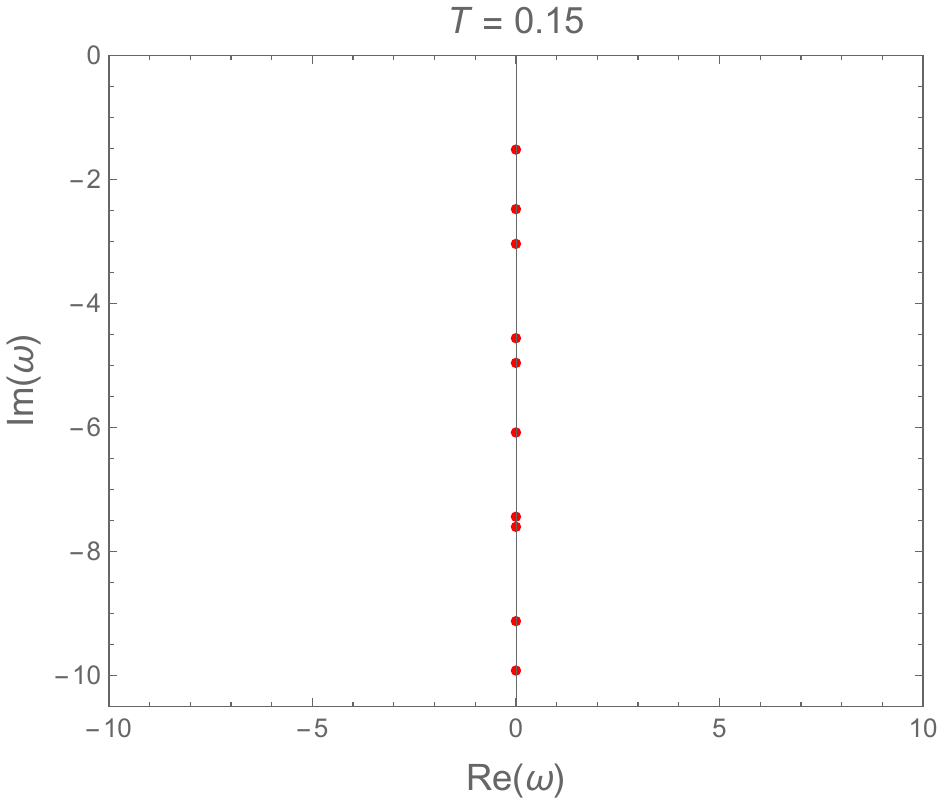}\qquad
  \includegraphics[width=0.3\textwidth]{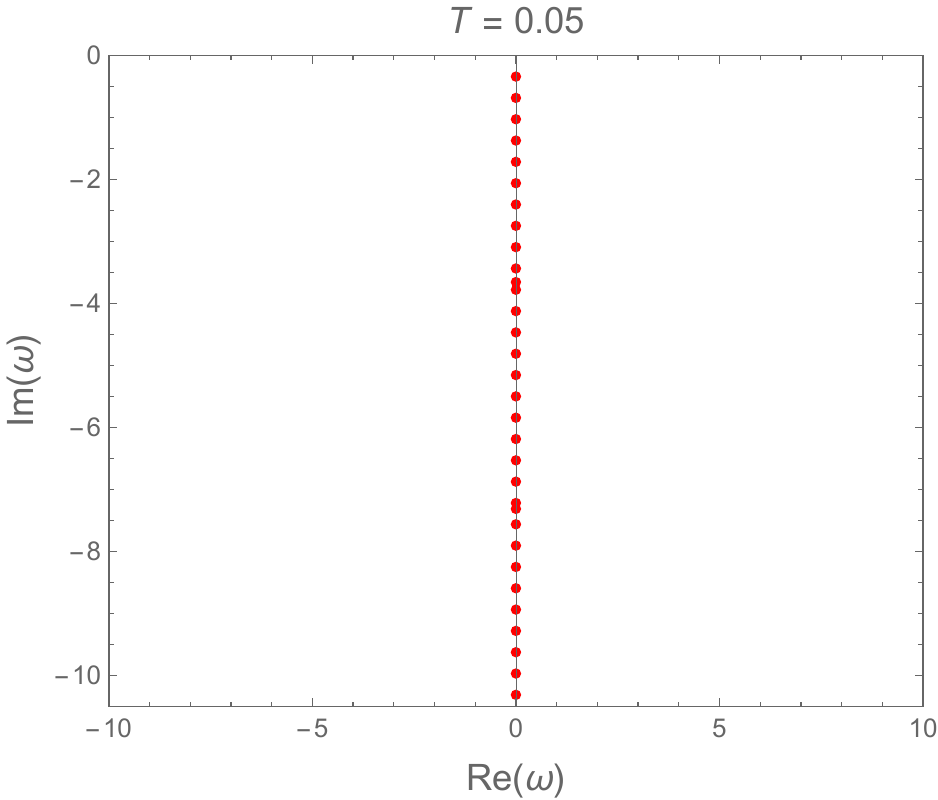}
  \caption{\label{fig:ads50} Quasinormal modes in the complex $\omega$-plane for planar Schwarzschild-AdS$_5$ black hole with momentum dissipation as the temperature decreases.}
\end{figure*}

When the axions vanish ($a=0$), the poles of the retarded Green's function are distributed below the real axis in the complex $\omega$-plane, where the solution can be found in Ref.~\cite{Kovtun:2005ev}. 
As the temperature decreases, i.e., $a$ increases, the poles move toward the negative imaginary axis. 
Once $a$ equals a critical value $a_0=2r_h/L$, the poles become constrained on the negative imaginary axis.
As the temperature is lowered, the poles become denser. More precisely, the poles which were on the right of the imaginary axis become sparser, and the poles which were on the left become denser.
The quasinormal frequencies as poles of the Green's function are shown in Fig.~\ref{fig:ads50}.

When the temperature is zero, $A_x(r)$ with the infalling boundary condition at the horizon can be solved by the Whittaker functions $W_{\kappa,\mu}(z)$:
\begin{equation}
  A_x=W_{\frac{i \omega}{4},\frac{1}{2}}\biggl(-\frac{i \omega}{r^2-1}\biggr).
\end{equation}
The Green's function is 
\begin{equation}
  G(\omega)=i\omega-\frac{1}{2} \omega ^2 \biggl[\psi \biggl(1-\frac{i \omega }{4}\biggr)+\log\left(-i\omega\right)+2\gamma\biggr].
\end{equation}
There are poles at $\omega_n=-4in$ ($n=1,2,3,\cdots$) and a branch cut due to the logarithmic term on the negative imaginary axis of the complex $\omega$-plane, which matches perfectly with the extremal limit of the finite temperature result.

\section{Conductivities in AdS$_4$}
\label{sec:ads4}
If we change the dimension to $d=3$ and perform the same calculation, we obtain that the conductivity is a constant. Without momentum dissipation, it is well known that the conductivity is a constant due to an electromagnetic self-duality \cite{Herzog:2007ij}. However, nonconstant conductivity in AdS$_4$ can be obtained in other systems.
We take the potential of the scalar field $\phi$ given by~\cite{Gao:2004tu}
\begin{multline}
  V(\phi)=-\frac{2}{(1+\alpha^2)^2L^2}\Bigl[\alpha^2(3\alpha^2-1)e^{-\phi/\alpha}\\
  +8\alpha^2e^{(\alpha-1/\alpha)\phi/2}+(3-\alpha^2)e^{\alpha\phi}\Bigr],\label{eq:potential4}
  \end{multline}
and $Z(\phi)=e^{-\alpha\phi}$,\footnote{When $Z=1$, the conductivity is a constant for arbitrary $\alpha$ and $a$.} where $\alpha$ is a parameter, and the values of $\alpha=1/\sqrt{3}$, $1$, and $\sqrt{3}$ correspond to special cases of STU supergravity. We call them 3-charge, 2-charge, and 1-charge black holes in AdS$_4$, respectively.\footnote{There are U$(1)^4$ gauge fields in STU supergravity in AdS$_4$ with charges $Q_{i}$. We call the system 3-charge black hole if $Q_1=Q_2=Q_3=Q$ and $Q_4=0$; 2-charge black hole if $Q_1=Q_2=Q$ and $Q_3=Q_4=0$; 1-charge black hole if $Q_1=Q$ and $Q_2=Q_3=Q_4=0$.} The system is invariant under $\alpha\to -\alpha$ and $\phi\to -\phi$, and we assume $\alpha>0$ in the following.

There is a neutral planar black hole solution for the system~\eqref{eq:axionsd} with the potential~\eqref{eq:potential4} \cite{Caldarelli:2016nni,Ren:2019lgw}. The factors $f(r)$ and $U(r)$ in the metric~\eqref{eq:metric} and the scalar field $\phi$ are
\begin{align}
  f(r) &=-\frac{a^2}{2}\biggl(1-\frac{b}{r}\biggr)^{\frac{1-\alpha^2}{1+\alpha^2}}+\frac{r^2}{L^2}\biggl(1-\frac{b}{r}\biggr)^{\frac{2\alpha^2}{1+\alpha^2}}\nonumber,\\
  U(r) &=r^2\biggl(1-\frac{b}{r}\biggr)^{\frac{2\alpha^2}{1+\alpha^2}},\quad e^{\alpha\phi}=\biggl(1-\frac{b}{r}\biggr)^\frac{2\alpha^2}{1+\alpha^2},\label{eq:sol4-axion}
\end{align}
where $a$ is associated with momentum dissipation, and $b$ is associated with the scalar field. The scaling dimension of the dual scalar operator is $\Delta_+=2$ or $\Delta_-=1$. The interpretation of the parameter $b$ depends on the boundary condition of the scalar field $\phi$ at the AdS boundary. For the Dirichlet boundary condition, $b$ is proportional to the source of the dual scalar operator in the standard quantization. A sourceless (mixed) boundary condition compatible with this solution is also applicable when a multi-trace deformation is present; $b$ is proportional to the expectation value of the dual scalar operator in the alternative quantization \cite{Caldarelli:2016nni}. Note that the parameter $b$ can be either positive or negative. For $\alpha\neq 0$, there are curvature singularities at $r=b$ and $r=0$. When $a\neq 0$, the curvature singularities are always enclosed by a horizon at $r=r_h$ at finite temperature.
Either $a$ or $b$ will be replaced by the black hole horizon size $r_h$ as needed.

First, we consider the case without momentum dissipation. When $a=0$, Eq.~\eqref{eq:sol4-axion} describes a spacetime without a regular horizon. Since the AdS boundary is at $r\to\infty$, the IR is at $r=b$ when $b>0$, and at $r=0$ when $b<0$. Moreover, the IR geometry is a hyperscaling violating geometry:
\begin{equation}
ds^2=\tilde{r}^\theta\biggl(-\frac{dt^2}{\tilde{r}^{2z}}+\frac{d\tilde{r}^2+d\vec{x}^2}{\tilde{r}^2}\biggr),
\end{equation}
with
\begin{equation}
z=1,\qquad \theta=\begin{cases}
\frac{2}{1-\alpha^2}\,\quad (b>0),
\\[2pt]
\frac{2\alpha^2}{\alpha^2-1}\,\quad (b<0).
\end{cases}
\end{equation}
For the background \eqref{eq:sol4-axion} with $a=0$, the perturbation equation~\eqref{eq:fluctuation} can be solved by modified Bessel functions as\footnote{The (modified) Bessel functions $Z_\nu(z)$ have branch points at $z=0$ and $z=\infty$, and the branch cut is conventionally chosen as the negative real axis.}
\begin{equation}
A_x=\sqrt{1-\frac{b}{r}}\,[C_1I_\nu(-ix)+C_2K_\nu(-ix)],
\label{eq:sol-ads4}
\end{equation}
where
\begin{equation}
\nu=\frac{1+\alpha ^2}{2 \left(1-\alpha ^2\right)},\qquad x=\frac{2\nu\omega L^2}{b}\biggl(1-\frac{b}{r}\biggr)^{\frac{1}{2\nu}}.
\end{equation}

It is crucial to examine the boundary conditions in the IR \cite{Charmousis:2010zz}. We assume $b>0$ first, in which case the IR is at $r=b$. When $0<\alpha<1$, there is no infalling wave, and we require that the solution is normalizable at $r=b$. Thus the solution is Eq.~\eqref{eq:sol-ads4} with $C_2=0$.\footnote{For certain values of $\alpha$, there are ambiguities for the boundary condition in the IR. This can be seen more clearly by writing Eq.~\eqref{eq:sol-ads4} as combinations of $J_\nu(x)$ and $J_{-\nu}(x)$. See
\cite{Charmousis:2010zz,Kiritsis:2015oxa}.}
When $b>0$ and $0<\alpha<1$, the Green's function is
\begin{equation}
   G(\omega)=
   i\omega\,\frac{I_{\nu-1}(-2i\nu\omega L^2/b)}{I_{\nu}(-2i\nu\omega L^2/b)}.
   \label{eq:G-I}
\end{equation}
The Green's function has poles on the real axis of the complex $\omega$-plane. When $\alpha>1$, we can impose infalling boundary condition in the IR, and it is more convenient to choose the modified Bessel function $K_\nu$ (or Hankel function $H_\nu^{(1)}$). When $b>0$ and $\alpha>1$, the Green's function is
  \begin{equation}
   G(\omega)=i\omega\,\frac{ K_{\nu-1}(2i\nu\omega L^2/b)}{K_{\nu}(2i\nu\omega L^2/b)}.
   \label{eq:G-K}
 \end{equation}
The Green's function has a branch cut on the negative imaginary axis of the complex $\omega$-plane. When $\alpha=1$, the result is Eq.~\eqref{eq:G-alpha-1} below. For all $\alpha>0$, the real part of the conductivity has a delta function at $\omega=0$, despite this is a zero density system.

For the case $b<0$, the IR geometry is at $r=0$. When $0<\alpha<1$, there is infalling wave in the IR, and the Green's function is Eq.~\eqref{eq:G-K}. When $\alpha>1$, the boundary condition in the IR is normalizability, and the Green's function is Eq.~\eqref{eq:G-I}. The conductivity no longer has a delta function at $\omega=0$. Depending on the parameters $b$ and $\alpha$, the above dilatonic model has rich features analogous to strongly correlated electronic systems, such as the twisted bilayer graphene \cite{Cao:2018}.

For a given $a\neq 0$, the system shares the same blackening factor as hyperbolic black holes, whose thermodynamics was studied in Ref.~\cite{Ren:2019lgw}.
We emphasize that the extremal limit of the $a\neq 0$ geometry may not be the $a=0$ geometry. To make this clear, consider the relation among $a$, $b$, and $r_h$ obtained by $f(r_h)=0$:
\begin{equation}
a^2=\frac{2}{L^2}r_h^\frac{3-\alpha^2}{1+\alpha^2}(r_h-b)^\frac{3\alpha^2-1}{1+\alpha^2}.
\end{equation}
The extremal geometry and its IR geometry is summarized as follows.
\begin{itemize}
\item $b>0$. The curvature singularity closer to the horizon is at $r=b$.
\begin{itemize}
\item[(i)] $\alpha>1/\sqrt{3}$. The extremal geometry has $a=0$ and $r_h=b$, and its IR is a hyperscaling-violating geometry.
\item[(ii)] $0<\alpha<1/\sqrt{3}$. The extremal geometry has $a\neq 0$ and $r_h>b$, and its IR is AdS$_2\times\mathbb{R}^2$.
\end{itemize}

\item $b<0$. The curvature singularity closer to the horizon is at $r=0$.
\begin{itemize}
\item[(i)] $0<\alpha<\sqrt{3}$. The extremal geometry has $a=0$ and $r_h=0$, and its IR is a hyperscaling-violating geometry.
\item[(ii)] $\alpha>\sqrt{3}$. The extremal geometry has $a\neq 0$ and $r_h>0$, and its IR is AdS$_2\times\mathbb{R}^2$.
\end{itemize}
\end{itemize}
The hyperscaling-violating geometry is the extremal limit of a regular black hole, if the Gubser criterion is satisfied \cite{Gubser:2000nd}. The values $\alpha=1/\sqrt{3}$ for $b>0$ and $\alpha=\sqrt{3}$ for $b<0$ are precisely the bound of the Gubser criterion.\footnote{In these two special cases, the extremal geometry has $a\neq 0$, and its IR is conformal to AdS$_2\times\mathbb{R}^2$.} Moreover, for a given $a\neq 0$, the black hole has a minimum temperature when $1/\sqrt{3}<\alpha<\sqrt{3}$.

When the momentum dissipation is nonzero, the solution~\eqref{eq:sol4-axion} describes finite temperature black holes. The DC conductivity is finite and can be calculated by the method in Refs.~\cite{Iqbal:2008by,Blake:2013bqa,Gouteraux:2014hca}:
\begin{equation}
\sigma_\text{DC}=Z[\phi(r_h)]=\biggl(1-\frac{b}{r_h}\biggr)^{-\frac{2\alpha^2}{1+\alpha^2}}.
\end{equation}
In the limit $a\to 0$, we have $\sigma_\text{DC}\to\infty$ for $b>0$, and $\sigma_\text{DC}=0$ for $b<0$. This matches the delta function in Eqs.~\eqref{eq:G-I} and \eqref{eq:G-K} at $\omega=0$ for $b>0$. In the following, we obtain analytic solutions of the holographic AC conductivity from the 3-charge, 2-charge, and 1-charge black holes in AdS$_4$ with momentum dissipation. We have checked that the $\omega\to 0$ limit of the AC conductivities below agrees with the DC result.

\subsection{3-charge black hole in AdS$_4$}
The solution for the 3-charge black hole in AdS$_4$ is given by Eq.~\eqref{eq:sol4-axion} with $\alpha=1/\sqrt{3}$. The temperature is
\begin{equation}
  T=\frac{\sqrt{r_h(r_h-b)}}{2\pi L^2},
  \label{eq:T-3-charge}
\end{equation}
where $r_h=aL/\sqrt{2}$. The solution of the perturbation equation~\eqref{eq:fluctuation} with the infalling boundary condition at the horizon is
\begin{equation}
A_x=\frac{r-b}{r+r_h}\biggl(\frac{r-r_h}{r+r_h}\biggr)^{-\frac{i\omega}{4\pi T}}{_2F_1}\biggl(\tilde{a},\tilde{b};\tilde{c};\tilde{x}\frac{r-r_h}{r+r_h}\biggr),
\end{equation}
where ${_2F_1}(\tilde{a},\tilde{b};\tilde{c};\tilde{x})$ is the hypergeometric function and
\begin{align}
&\tilde{a}=1-\frac{i\omega L^2}{2} \biggl(\frac{1}{\sqrt{r_h (r_h-b)}}-\frac{1}{\sqrt{r_h (r_h+b)}}\biggr),\nonumber\\
&\tilde{b}=1-\frac{i\omega L^2}{2} \biggl(\frac{1}{\sqrt{r_h (r_h-b)}}+\frac{1}{\sqrt{r_h (r_h+b)}}\biggr),\nonumber\\
&\tilde{c}=1-\frac{i\omega L^2}{\sqrt{r_h(r_h-b)}},\qquad \tilde{x}=\frac{b+r_h}{b-r_h}.
\end{align}
By the AdS/CFT prescription, the Green's function is given by
\begin{equation}
 G(\omega)=-\frac{b}{L^2}-\frac{r_h}{L^2}\biggl(\tilde{c}+\frac{2\tilde{a}\tilde{b}\tilde{x}}{\tilde{c}}\cdot\frac{{_2F_1}(1+\tilde{a},1+\tilde{b};1+\tilde{c};\tilde{x})}{{_2F_1}(\tilde{a},\tilde{b};\tilde{c};\tilde{x})}\biggr).\label{eq:greenads4-3}
\end{equation}

We need to treat the zero temperature limit more carefully, since the $b<0$ and $b>0$ cases have different extremal geometries. When $b<0$, the extremal limit is at $r_h=0$, in which case the axions vanish ($a=0$). The conductivity is given by Eq.~\eqref{eq:G-K}. When $b>0$, the extremal limit is at $r_h=b$, in which case the axions do not vanish. For the $b>0$ extremal geometry, we solve the perturbation equation~\eqref{eq:fluctuation} and obtain the Green's function at $T=0$. After imposing the infalling boundary condition in the IR, the solution of $A_x(r)$ is
\begin{equation}
  A_x= K_{\tilde{\nu}}\biggl(-\tilde{\nu}\sqrt{\frac{r+b}{r-b}}\,\biggr),\qquad \tilde{\nu}=\frac{i\omega L^2}{\sqrt{2}b}.
\end{equation}
The Green's function is
\begin{equation}
  G\left( \omega \right)=\frac{i\omega}{\sqrt{2}}\biggl(\frac{K_{\tilde{\nu}+1}(-\tilde{\nu})}{K_{\tilde{\nu}}(-\tilde{\nu})}+1\biggr).
\end{equation}
The Green's function has a branch cut on the negative imaginary axis of the complex $\omega$-plane.

\subsection{2-charge black hole in AdS$_4$}
The solution for the 2-charge black hole in AdS$_4$ is given by Eq.~\eqref{eq:sol4-axion} with $\alpha=1$. 
The temperature is
\begin{equation}
  T=\frac{2r_h-b}{4\pi L^2}=\frac{2r_h^2+a^2L^2}{8\pi r_hL^2},
\end{equation}
where $a^2L^2=2r_h(r_h-b)$. There is a minimum temperature. The extremal limit is always at $a=0$, in contrast to the 3-charge black hole in AdS$_4$.  The extremal geometry apparently has a finite temperature despite there is no horizon enclosing the spacetime singularity. The solution of the perturbation equation~\eqref{eq:fluctuation} with the infalling boundary condition at the horizon is
\begin{multline}
A_x=
\biggl(\frac{r-r_h}{r+r_h-b}\biggr)^{-\frac{i\omega }{4\pi T}}H\ell\biggl(\frac{\left(2 r_h-b\right) b}{r_h^2},\\
-\frac{i\omega L^2b}{r_h^2};\, 0,\, 2,\, 1-\frac{2 i\omega L^2}{2 r_h-b},\, 1;\, \frac{b(r-r_h)}{r_h (r-b)}\biggr),\label{eq:ads42ax}
\end{multline}
where $H\ell(\tilde{a},\tilde{q};\tilde{\alpha},\tilde{\beta},\tilde{\gamma},\tilde{\delta};\tilde{z})$ is the (local) Heun function. The Green's function is given by
\begin{align}
&G(\omega)=i\omega -\frac{(r_h-b)b}{r_hL^2}\times\nonumber\\
&\,\times\frac{H\ell'\Bigl(\frac{(2r_h-b)b}{r_h^2},\, -\frac{i\omega L^2b}{r_h^2};\, 0,\, 2,\, 1-\frac{2 i\omega L^2}{2r_h-b},\, 1;\, \frac{b}{r_h}\Bigr)}{H\ell\Bigl(\frac{(2r_h-b)b}{r_h^2},\, -\frac{i\omega L^2b}{r_h^2};\, 0,\, 2,\, 1-\frac{2 i\omega L^2}{2r_h-b},\, 1;\, \frac{b}{r_h}\Bigr)}.
\label{eq:G-ads42}
\end{align}
where $H\ell'(\tilde{a},\tilde{q};\tilde{\alpha},\tilde{\beta},\tilde{\gamma},\tilde{\delta};\tilde{z})$ is the derivative of the Heun function.

\begin{figure*}
  \includegraphics[width=0.3\textwidth]{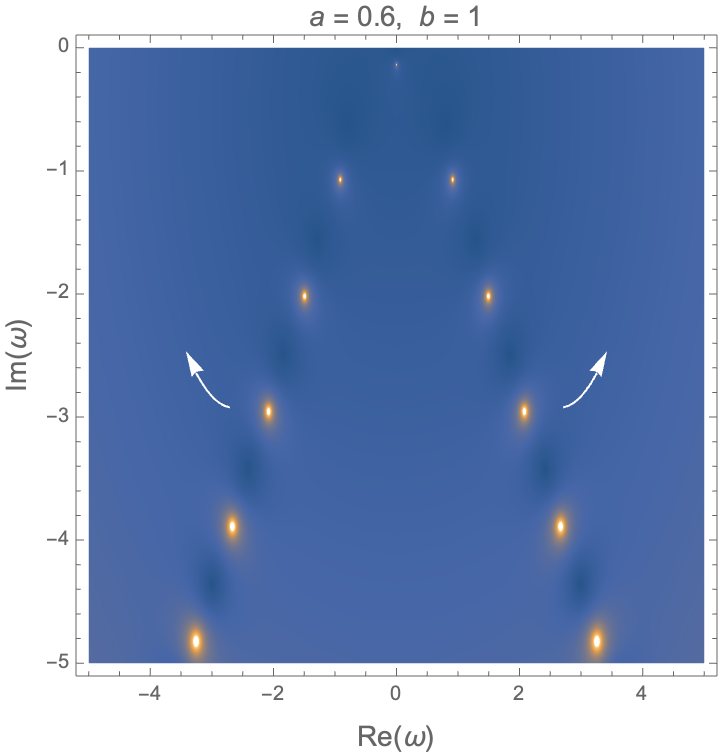}\qquad
  \includegraphics[width=0.3\textwidth]{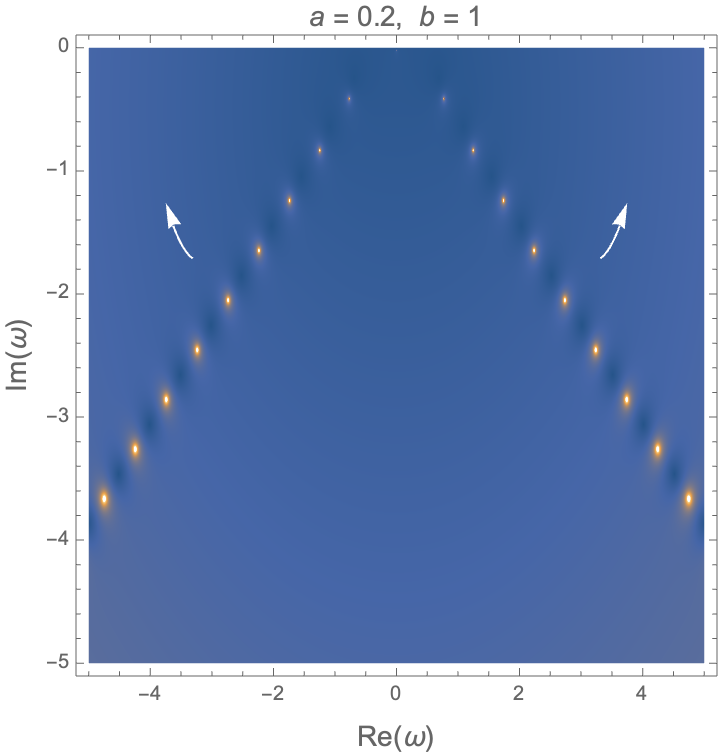}\\
  \caption{\label{fig:ads42} Quasinormal modes in the complex $\omega$-plane for the 2-charge black hole in AdS$_4$ with momentum dissipation. The arrows illustrate the moving of poles as the extremal limit ($a=0$) is approached.}
\end{figure*}

The conductivity without momentum dissipation has been obtained in Ref.~\cite{Kiritsis:2015oxa} for $b>0$. The following result is for general $b$. When the axions vanish ($a=0$), the solution of $A_x(r)$ is
\begin{equation}
A_x=C_1\biggl(\frac{r-b}{r}\biggr)^\frac{b+\sqrt{b^2-4\omega^2L^4}}{2b}+C_2\biggl(\frac{r-b}{r}\biggr)^\frac{b-\sqrt{b^2-4\omega^2L^4}}{2b}.
\end{equation}
When $2\omega L^2>|b|$, the first term with the $\omega\to\omega+i\epsilon$ prescription describes infalling wave in the IR for either $b>0$ or $b<0$. Thus the infalling boundary condition requires $C_2=0$. The Green's function is given by
\begin{equation}
  G(\omega)=-\frac{1}{2}\Bigl(b+\sqrt{b^2-4\omega^2L^4}\Bigr).
\label{eq:G-alpha-1}
\end{equation}
There are branch cuts $4\omega^2L^4>b^2$ on the real axis of the complex $\omega$-plane. For small $a$, the branch cuts become dense poles in the lower-half plane. As $a$ increases, the poles become increasingly sparser and closer to the imaginary axis of the complex $\omega$-plane. The quasinormal frequencies as poles of the Green's function are shown in Fig.~\ref{fig:ads42}.

\begin{figure}
  \includegraphics[width=0.39\textwidth]{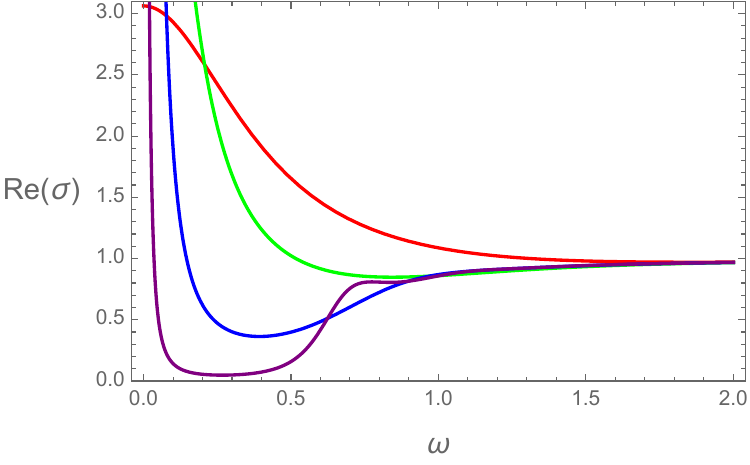}
  \caption{\label{fig:sigma} Real part of the conductivity as a function of frequency plotted  for the 2-charge black hole in AdS$_4$. We take $b=1$ and $L=1$. The red, green, blue, and purple curves are for $a=1.2$, $0.6$, $0.2$, and $0.05$, respectively.}
\end{figure}

The conductivity is obtained by $\sigma=G/i\omega|_{\omega\to\omega+i\epsilon}$, and the real part is
\begin{equation}
\text{Re}[\sigma(\omega)]=\frac{\pi b}{L^2}\,\theta(b)\delta(\omega)+\theta(4\omega^2L^4-b^2)\frac{\sqrt{4\omega^2L^4-b^2}}{2\omega L^2},
\label{eq:sigma4-2}
\end{equation}
where $\theta(x)$ is the Heaviside step function. There is a delta function at $\omega=0$ when $b>0$, and there is a gap in $0<2\omega L^2<|b|$. When $a$ is close to zero, we expect that the delta function becomes a Drude-like peak. Note that the system is at zero density. It was pointed out that this type of delta function at $\omega=0$ together with the hard gap is closely related to superconductivity \cite{DeWolfe:2012uv}, and an example was given in Ref.~\cite{Aprile:2012sr}. Figure~\ref{fig:sigma} shows the real part of the conductivity calculated from Eq.~\eqref{eq:G-ads42} as the extremal limit is approached.

\subsection{1-charge black hole in AdS$_4$}
The solution for the 1-charge black hole in AdS$_4$ is given by Eq.~\eqref{eq:sol4-axion} with $\alpha=\sqrt{3}$. The temperature is
\begin{equation}
  T=\frac{\sqrt{r_h(r_h-b)}}{2\pi L^2},
  \label{eq:T-1-charge}
\end{equation}
where $r_h=aL/\sqrt{2}+b$. The solution of the perturbation equation~\eqref{eq:fluctuation} with the infalling boundary condition at the horizon is
\begin{multline}
A_x=\frac{r+r_h-2b}{r}\biggl(\frac{r-r_h}{r+r_h-2b}\biggr)^{-\frac{i\omega}{4\pi T}}\\
\times H\ell\biggl(\tilde{a},\, \tilde{q};\,\tilde{\alpha},\, \tilde{\beta},\, \tilde{\gamma},\, \tilde{\delta};\, \tilde{z}\frac{r-r_h}{r+r_h-2b}\biggr),\label{eq:ads41ax}
\end{multline}
where $H\ell$ is the Heun function and
\begin{align}
&\tilde{a}=1-\frac{2b}{r_h},\quad
\tilde{q}=-\tilde{a}+\frac{i\omega L^2}{\sqrt{r_h(r_h-b)}}+\frac{\omega^2L^4b}{2r_h^2(r_h-b)},\nonumber\\
&\tilde{\alpha}=-1-\frac{i\omega L^2}{2} \biggl(\frac{1}{\sqrt{r_h (r_h-b)}}-\frac{1}{\sqrt{(r_h-b)(r_h-2b)}}\biggr),\nonumber\\
&\tilde{\beta}=-1-\frac{i\omega L^2}{2} \biggl(\frac{1}{\sqrt{r_h (r_h-b)}}+\frac{1}{\sqrt{(r_h-b)(r_h-2b)}}\biggr),\nonumber\\
&\tilde{\gamma}=1-\frac{i\omega L^2}{\sqrt{r_h(r_h-b)}},\quad \tilde{\delta}=0\,,\quad
\tilde{z}=-\tilde{a}\,.
\end{align}
By the AdS/CFT prescription, the Green's function is given by
\begin{align}
&G(\omega)=i\omega\sqrt{\tilde{b}}+\frac{r_h\tilde{a}}{L^2}\biggl(1+2\tilde{b}\,\frac{H\ell'\bigl(\tilde{a},\, \tilde{q};\,\tilde{\alpha},\, \tilde{\beta},\, \tilde{\gamma},\, \tilde{\delta};\,\tilde{z}\bigr)}{H\ell\bigl(\tilde{a},\, \tilde{q};\,\tilde{\alpha},\, \tilde{\beta},\, \tilde{\gamma},\, \tilde{\delta};\,\tilde{z}\bigr)}\biggr),
\label{eq:G-ads41}
\end{align}
where $\tilde{b}\equiv 1-b/r_h$, and $H\ell'$ is the derivative of the Heun function.

We need to treat the zero temperature limit more carefully, since the $b<0$ and $b>0$ cases have different extremal geometries. When $b>0$, the extremal limit is at $r_h=b$, in which case the axions vanish ($a=0$). The conductivity is given by Eq.~\eqref{eq:G-K}. When $b<0$, the extremal limit is at $r_h=0$, in which case the axions do not vanish. For the $b<0$ extremal geometry, the perturbation equation~\eqref{eq:fluctuation} can be solve in terms of a confluent Heun function.

\section{Conductivities in AdS$_5$}
\label{sec:ads5}
The AdS$_5$ counterpart of the preceding section is as follows.
We take the potential of the scalar field $\phi$ given by \cite{Gao:2004tv}
\begin{multline}
V(\phi)=-\frac{12}{(4+3\alpha^2)^2L^2}\Bigl[3\alpha^2(3\alpha^2-2)e^{-\frac{4\phi}{3\alpha}}\\
+36\alpha^2e^{\frac{3\alpha^2-4}{6\alpha}\phi}+2(8-3\alpha^2)e^{\alpha\phi}\Bigr],
\label{eq:potential5}
\end{multline}
and $Z(\phi)=e^{-\alpha\phi}$, where $\alpha$ is a parameter, and the values of $\alpha=2/\sqrt{6}$ and $4/\sqrt{6}$ correspond to special cases of STU supergravity. We call them 2-charge and 1-charge black holes in AdS$_5$, respectively.\footnote{There are U$(1)^3$ gauge fields in STU supergravity in AdS$_5$ with charges $Q_{i}$. We call the system 2-charge black hole if $Q_1=Q_2=Q$ and $Q_3=0$; 1-charge black hole if $Q_1=Q$ and $Q_2=Q_3=0$.}

There is a neutral planar black hole solution for the system~\eqref{eq:axionsd} with the potential~\eqref{eq:potential5}. The solution of the metric $g_{\mu\nu}$ and the scalar field $\phi$ is
\begin{align}
& ds^2 =-f(r) dt^2+g(r)^{-1} dr^2+U(r) d\mathbf{x}^2\,,\label{eq:fsol5-1}\\
& e^{\alpha\phi} =\left(1-\frac{b^2}{r^2}\right)^\frac{6\alpha^2}{4+3\alpha^2},
\end{align}
with
\begin{equation}
\begin{split}
f &=-\frac{a^2}{4}\left(1-\frac{b^2}{r^2}\right)^\frac{4-3\alpha^2}{4+3\alpha^2}+\frac{r^2}{L^2}\left(1-\frac{b^2}{r^2}\right)^\frac{3\alpha^2}{4+3\alpha^2},\\
g &=f(r)\left(1-\frac{b^2}{r^2}\right)^{\frac{3\alpha^2}{4+3\alpha^2}},\qquad
U=r^2\left(1-\frac{b^2}{r^2}\right)^\frac{3\alpha^2}{4+3\alpha^2},
\end{split}\label{eq:fsol5-2}
\end{equation}
where the parameter $b^2$ ($b>0$) can be either positive or negative. For $\alpha\neq 0$, there are curvature singularities at $r^2=b^2>0$ and $r=0$. When $a\neq 0$, the curvature singularities are always enclosed by a horizon at finite temperature. When $a=0$, Eq.~\eqref{eq:fsol5-2} describes a spacetime without a regular horizon. Since the AdS boundary is at $r\to\infty$, the IR is at $r=b$ when $b^2>0$, and at $r=0$ when $b^2<0$.

Much less analytic solutions of the AC conductivity are available in AdS$_5$ than in AdS$_4$. We can obtain analytic solutions in two special cases without momentum dissipation, which are 2-charge black hole in AdS$_5$ with $b^2<0$ and 1-charge black hole in AdS$_5$ with $b^2>0$. The solution for the later case has been found in Ref.~\cite{DeWolfe:2012uv}.

\textit{2-charge black hole in AdS$_5$} ($\alpha=2/\sqrt{6}$) \textit{with} $b^2<0$. We define $\bar{b}^2=-b^2>0$ ($\bar{b}>0$), and the curvature singularity is at $r=\bar{b}$.
When the axions vanish ($a=0$), the solution of $A_x(r)$ is
\begin{multline}
A_x=\biggl(\frac{r}{\bar{b}}\biggr)^{-1+\frac{\sqrt{\bar{b}^2-\omega ^2L^4}}{\bar{b}}} \, _2F_1\biggl(\frac{1}{2}+\frac{\sqrt{\bar{b}^2-\omega ^2L^4}}{2 \bar{b}},\\
-\frac{1}{2}+\frac{\sqrt{\bar{b}^2-\omega^2L^4}}{2 \bar{b}};1+\frac{\sqrt{\bar{b}^2- \omega^2L^4}}{\bar{b}};-\frac{r^2}{\bar{b}^2}\biggr).
\end{multline}
When $\omega L^2>\bar{b}$, the above solution with the $\omega\to\omega+i\epsilon$ prescription describes infalling wave in the IR. The Green's function is given by
\begin{equation}
G(\omega)=-\omega^2\biggl[\psi\biggl(\frac{1}{2}+\frac{\sqrt{\bar{b}^2-\omega^2L^4}}{2 \bar{b}}\biggr)+\log\bar{b}+\gamma\biggr].
\label{eq:G5-2}
\end{equation}
There are branch cuts $\omega^2L^4>b^2$ on the real axis of the complex $\omega$-plane.

\textit{1-charge black hole in AdS$_5$} ($\alpha=4/\sqrt{6}$) \textit{with} $b^2>0$. The IR is at $r=b$. It is more convenient to define $\bar{r}^2=r^2-b^2$ so that the IR is at $\bar{r}=0$. When the axions vanish ($a=0$), the solution of $A_x(\bar{r})$ is
\begin{multline}
A_x=\biggl(\frac{\bar{r}}{b}\biggr)^{1+\frac{\sqrt{b^2-\omega^2L^4}}{b}} \, _2F_1\biggl(\frac{1}{2}+\frac{\sqrt{b^2-\omega^2L^4}}{2 b},\\
\frac{3}{2}+\frac{\sqrt{b^2-\omega^2L^4}}{2 b};1+\frac{\sqrt{b^2-\omega^2L^4}}{b};-\frac{\bar{r}^2}{b^2}\biggr).
\end{multline}
When $\omega L^2>b$, the above solution with the $\omega\to\omega+i\epsilon$ prescription describes infalling wave in the IR. The Green's function is given by
\begin{equation}
G(\omega)=-\frac{2b^2}{L^4}-\omega^2\biggl[\psi\biggl(\frac{1}{2}+\frac{\sqrt{b^2- \omega^2L^4}}{2 b}\biggr)+\log b+\gamma\biggr].
\label{eq:G5-1}
\end{equation}
There are branch cuts $\omega^2L^4>b^2$ on the real axis of the complex $\omega$-plane. The only difference between Eqs.~\eqref{eq:G5-1} and \eqref{eq:G5-2} is that Eq.~\eqref{eq:G5-1} has an extra term that gives a delta function at $\omega=0$ in the real part of the conductivity.

The conductivity is obtained by $\sigma=G/i\omega|_{\omega\to\omega+i\epsilon}$, and the real part is 
\begin{multline}
\text{Re}[\sigma(\omega)]=\frac{2\pi b^2}{L^4}\delta(\omega)\\
+\frac{\pi\omega}{2}\theta(\omega^2L^4-b^2)\tanh\frac{\pi\sqrt{\omega^2L^4-b^2}}{2b}\,,
\end{multline}
where $\theta(x)$ is the Heaviside step function. There is a delta function at $\omega=0$, and there is a gap at $0<\omega L^2<b$. This result was in Ref.~\cite{DeWolfe:2012uv}. The AdS$_4$ counterpart of this result is Eq.~\eqref{eq:sigma4-2}. When $a$ is close to zero, we expect that the delta function becomes a Drude-like peak.

For a given $a\neq 0$, the system shares the same blackening factor as hyperbolic black holes, whose thermodynamics was studied in Ref.~\cite{Ren:2019lgw}.
We emphasize that the extremal limit of the $a\neq 0$ geometry may not be the $a=0$ geometry. To make this clear, consider the relation among $a$, $b$, and $r_h$ obtained by $f(r_h)=0$:
\begin{equation}
a^2=\frac{4}{L^2}(r_h^2)^{\frac{8-3\alpha^2}{4+3\alpha^2}}(r_h^2-b^2)^\frac{6\alpha^2-4}{4+3\alpha^2}.
\end{equation}
The extremal geometry and its IR geometry is summarized as follows.
\begin{itemize}
\item $b^2>0$. The curvature singularity closer to the horizon is at $r=b$.
\begin{itemize}
\item[(i)] $\alpha>2/\sqrt{6}$. The extremal geometry has $a=0$ and $r_h=b$, and its IR is a hyperscaling-violating geometry.
\item[(ii)] $0<\alpha<2/\sqrt{6}$. The extremal geometry has $a\neq 0$ and $r_h>b$, and its IR is AdS$_2\times\mathbb{R}^3$.
\end{itemize}

\item $b^2<0$. The curvature singularity closer to the horizon is at $r=0$.
\begin{itemize}
\item[(i)] $0<\alpha<4/\sqrt{6}$. The extremal geometry has $a=0$ and $r_h=0$, and its IR is a hyperscaling-violating geometry.
\item[(ii)] $\alpha>4/\sqrt{6}$. The extremal geometry has $a\neq 0$ and $r_h>0$, and its IR is AdS$_2\times\mathbb{R}^3$.
\end{itemize}
\end{itemize}
The hyperscaling-violating geometry is the extremal limit of a regular black hole, if the Gubser criterion is satisfied. The values $\alpha=2/\sqrt{6}$ for $b^2>0$ and $\alpha=4/\sqrt{6}$ for $b^2<0$ are precisely the bound of the Gubser criterion. Moreover, for a given $a\neq 0$, the black hole has a minimum temperature when $2/\sqrt{6}<\alpha<4/\sqrt{6}$.

\section{R-current correlator of $\mathcal{N}=4$ SYM on a sphere}
\label{sec:R-current}
Although the planar black holes with momentum dissipation are from the bottom-up approach, they are closely related to spherical black holes in the top-down approach. The blackening factor for a hyperbolic black hole is the same as a planar black hole with momentum dissipation \cite{Bardoux:2012aw,Gouteraux:2014hca,Ren:2019lgw}. By analytic continuation of the solution for a hyperbolic black hole, we can obtain the solution for a spherical black hole.

An R-current correlator of $\mathcal{N}=4$ SYM on a sphere at finite temperature in the large $N$ and strong coupling limit is obtained as follows, as a generalization of the previous result in which the CFT lives on flat space \cite{Kovtun:2005ev,Myers:2007we}. The metric for the spherical Schwarzschild-AdS$_5$ black hole is
\begin{equation}
  ds^2=-f(r)\tsp dt^2+f(r)^{-1}\tsp dr^2+r^2\tsp d\Omega_3^2\,,\label{eq:metric-sph}
\end{equation}
where $d\Omega_3^2=d\rho^2+{\sin^2\rho}\,(d\theta^2+\sin^2\theta\,d\varphi^2)$ is the metric for a 3-dimensional sphere with unit radius. The solution of $f(r)$ is Eq.~\eqref{eq:fsol} with $a^2=-4$.\footnote{By setting $a^2=4$, we can obtain a hyperbolic black hole; the boundary theory lives in a hyperbolic space with radius $L$.} The boundary CFT lives on a sphere with radius $L$. The temperature is given by
\begin{equation}
  T=\frac{2r_h^2+L^2}{2\pi r_h L^2}.
\end{equation}
The black hole has a minimum temperature at $r_h=L/\sqrt{2}$. There is a Hawking-Page phase transition between the large black hole and thermal AdS \cite{Hawking:1982dh,Witten:1998zw}.

We perturb the system by\footnote{The same perturbation equation can be obtained by $\delta A_\rho=e^{-i\omega t}A_\rho(r)\csc^2\rho$ or $\delta A_\theta=e^{-i\omega t}A_\theta(r)\csc\theta$.}
\begin{equation}
\delta A_\varphi=e^{-i\omega t}A_\varphi(r)\,.
\end{equation}
The Maxwell equation is exactly the same as Eq.~\eqref{eq:fluctuation}, provided that we use the new $f(r)$ for the spherical geometry. The solution with the infalling boundary condition at the horizon is
\begin{align}
A_\varphi=\biggl(\frac{r^2-r_h^2}{r^2+r_h^2+L^2}\biggr)^{-\frac{i\omega}{4\pi T}} \,{_2F_1}\biggl(\frac{\omega L^2 (\sqrt{r_h^2+L^2}-i r_h)}{2 (2 r_h^2+L^2)},\nonumber\\
-\frac{\omega L^2 (\sqrt{r_h^2+L^2}+i r_h)}{2 (2 r_h^2+L^2)};1-\frac{i\omega L^2 r_h}{2 r_h^2+L^2};\frac{r^2-r_h^2}{r^2+r_h^2+L^2}\biggr).
\end{align}
By the AdS/CFT prescription, the Green's function is obtained as
\begin{align}
& G(\omega)=-i\omega \frac{r_h}{L^2}-\frac{1}{2}\omega^2 \biggl[\psi\biggl(\frac{\omega L^2 (\sqrt{r_h^2+L^2}-i r_h)}{2 \left(2 r_h^2+L^2\right)}\biggr)\nonumber\\
& \;+\psi\biggl(-\frac{\omega L^2 (\sqrt{r_h^2+L^2}+i r_h)}{2 \left(2 r_h^2+L^2\right)}\biggr)+2 \gamma +\log \left(2 r_h^2+L^2\right)\biggr].
\end{align}
The poles of the Green's function are at
\begin{equation}
\omega_nL^2=2n\Bigl(\pm\sqrt{r_h^2+L^2}-ir_h\Bigr),
\end{equation}
where $n=1,2,3,\cdots$.

\section{Discussion}
\label{sec:sum}
We have obtained analytic solutions of the AC conductivity at arbitrary frequency and temperature from the planar Schwarzschild-AdS$_5$ black hole and dilatonic black holes with momentum dissipation. In the extremal limit, these black hole solutions have distinctive IR geometries characterizing low energy properties of the dual condensed matter systems. With the analytic solutions in hand, we study analyticity properties of the Green's function, especially how the poles move in the complex $\omega$-plane move as the temperature varies. Typically, branch cuts emerge when the extremal limit is approached. At zero temperature, the real part of the conductivity can have a delta function at $\omega=0$ and a hard gap. The ground states have rich properties, depending on the parameters $a$ (momentum dissipation), $b$ (scalar field), and $\alpha$ (potential). Ground state properties of the AdS$_4$ dilatonic system are summarized in Table~\ref{tab:GS}.
\begin{table}
\caption{\label{tab:GS} Ground state properties of the AdS$_4$ dilatonic system.}
\renewcommand{\arraystretch}{1.5}
\setlength\doublerulesep{0.05pt}
\begin{ruledtabular}
\begin{tabular}{ccccc}
$b$ & $\alpha$ & $a$ & $\sigma_\text{DC}$ & $\sigma(\omega)$\\
\hline
& $0<\alpha<1/\sqrt{3}$ & $a\neq 0$ & finite & gapless\\
$b>0$ & $1/\sqrt{3}<\alpha<1$ & $a=0$ & $\infty$ & gapped\\
& $\alpha>1$ & $a=0$ & $\infty$ & gapless\\
\hline
& $0<\alpha<1$ & $a=0$ & 0 & gapless\\
$b<0$ & $1<\alpha<\sqrt{3}$ & $a=0$ & 0 & gapped\\
& $\alpha>\sqrt{3}$ & $a\neq 0$ & finite & gapless\\
\end{tabular}
\end{ruledtabular}
\end{table}

A number of aspects need more study in the future.
(i) We have not explored the full parameter space for the dilatonic models in AdS$_4$ and AdS$_5$. It may shed some light on understanding strongly correlated materials.
(ii) The 2-charge black hole in AdS$_5$ and the 3-charge black hole in AdS$_4$ are also called the Gubser-Rocha model \cite{Gubser:2009qt} and have been intensely studied. Other black holes in STU supergravity also have distinctive and interesting properties.
(iii) How does the strongly coupled result relate to the weakly coupled result as we tune the coupling constant \cite{Hartnoll:2005ju,Grozdanov:2018gfx}? It would be interesting to make a comparison between the correlators calculated by AdS/CFT and field theory techniques.

\begin{acknowledgments}
J.R. thanks Christopher Herzog, Elias Kiritsis, and Li Li for helpful discussions. This work was supported in part by the NSF of China under Grant No. 11905298.
\end{acknowledgments}

\nocite{*}


\end{document}